\documentclass[useAMS,usenatbib]{mn2e}
\usepackage{graphicx}

\def\be{\begin{equation}}
\def\ee{\end{equation}}
\def\ba{\begin{eqnarray}}
\def\ea{\end{eqnarray}}

\def\ltsima{$\; \buildrel < \over \sim \;$}
\def\simlt{\lower.5ex\hbox{\ltsima}}
\def\gtsima{$\; \buildrel > \over \sim \;$}
\def\simgt{\lower.5ex\hbox{\gtsima}}
\def\etal{{et al.\ }}

\usepackage[dvipsnames]{color}

\title[Thermal instability in the collisionally cooled gas]
{Thermal instability in the collisionally cooled gas}
\author[E. O. Vasiliev]
       {Evgenii O. Vasiliev$^{1,2}$\thanks{E-mail:eugstar@mail.ru}\\
$^1$Institute of Physics, Southern Federal University, Stachki Ave. 194, Rostov-on-Don, 344090 Russia\\
$^2$Institute of Astronomy, Russian Academy of Sciences, Pyatnitskaya st. 48, Moscow 119017 Russia\\
}
\begin{document}
\date{Accepted 3004 December 15.
      Received 2004 December 14;
      in original form 2004 December 31}
\pagerange{\pageref{firstpage}--\pageref{lastpage}}
\pubyear{3004}
\maketitle

\label{firstpage}

\begin{abstract}
We have presented the non-equilibrium (time-dependent) cooling rate and ionization state calculations for
a gas behind shock waves with $v \sim 50-150$~km~s$^{-1}$ ($T_s \sim 0.5 - 6\times 10^5$~K). Such shock 
waves do not lead to the radiative precursor formation, i.e. the thermal evolution of a gas behind the shock
waves are controlled by collisions only. We have found that the cooling rate in a gas behind the shock waves 
with $v \sim 50-120$~km~s$^{-1}$ ($T_s \sim 0.5 - 3\times 10^5$~K) differs considerably from the cooling rate 
for a gas cooled from $T = 10^8$~K. It is well-known that a gas cooled from $T = 10^8$~K is thermally unstable 
for isobaric and isochoric perturbations at $T \simgt 2\times 10^4$~K. We have studied the thermal instability 
in a collisionally controlled gas for shock waves with $v \sim 50-150$~km~s$^{-1}$. We have found that the 
temperature range, where the postshock gas is thermally unstable, is significantly modified and depends on both 
gas metallicity and ionic composition of a gas before shock wave. For $Z \simgt 0.1Z_\odot$ the temperature 
range, where the thermal instability criterion for isochoric perturbations is not fulfilled, widens in comparison 
with that for a gas cooled from $T = 10^8$~K, while that for isobaric perturbations remains almost without a change. 
For $Z\sim Z_\odot$ a gas behind shock waves with $v \simlt 65$~km~s$^{-1}$ ($T_s \simlt 10^5$~K) is thermally 
stable to isochoric perturbations during full its evolution. We have shown that the transition from isobaric 
to isochoric cooling for a gas with $Z \simgt 0.1Z_\odot$ behind shock waves with $T_s = 0.5 - 3\times 10^5$~K 
proceeds at lower column density layer behind a shock wave than that for a gas cooled from $T = 10^8$~K. The ionic
states in a gas with $Z \sim 10^{-3}-1~Z_\odot$ behind shock waves with $T_s \simlt 4\times 10^5$~K demostrate a 
significant difference from these in a gas cooled from $T = 10^8$~K. Such difference is thought to be important 
for correct interpretation of observational data, but hardly help to dicriminate thermally stable gas.
\end{abstract}


%----------------------- Section 1 -------------------------------

\section{Introduction}

\noindent

Thermal instability (TI) comprehensively analyzed by \citet{field} is frequently considered as a 
good candidate to explain the planetary nebulae evolution \citep{hunter},
cooling flows \citep{nulsen}, the formation of the interstellar clouds \citep{burkert}, 
the temperature distribution of the unstable interstellar gas 
in the Galactic disk \citep{scalo}, the formation of high velocity clouds in galactic corona 
\citep{binney} and so on. The most interesting question is a possibilty
of the TI development behind shock waves. The analysis of stability in the postshock flow was performed 
by \citet{yua78,yua79} and \citet{chevalier}. Further progress is reached by \citet{nishi}, 
who have analyzed the thermal instability behind radiative shock waves and studied its possible 
role in the fragmentation. In the numerical simulations the role of thermal instability behind radiative 
shock waves was studied by many authors \citep[e.g.,][]{dopita85,bregman,sd03}. 
In the above-metioned papers the radiative shock waves with velocities 
$\sim$150 -- 1500~km~s$^{-1}$ are considered. On the other hand the thermal/thermo-reactive 
instabilities are analyzed in a diffuse interstellar photoionized gas with temperature below $10^4$~K
\citep{ferrara95,smithTI}.
Shock waves with velocities $\sim 50-150$~km~s$^{-1}$ are quite important for the evolution of dwarf 
galaxies. Such shock velocities can be assotiated with global star formation processes and winds in 
dwarf galaxies.  Thus, it is interesting to study both a possibility
of the thermal instability and the ionization and thermal evolution in a gas behind shock waves with
$v \sim 50-150$~km~s$^{-1}$.

Such shock waves are belived to play a significant role in the metal enrichment of the intergalactic 
medium (IGM). Metals (heavy elements) produced by stars are transported into the IGM by shock waves 
from galaxies and clusters of galaxies \citep{gnedin,madau01,yua02,schayeigm,meiksin}. The efficiency of 
metal ejection depends on many factors and parameters, e.g. mass of a parent (for metals) galaxy, star 
formation rate, density profile etc. On one hand massive galaxies should produce and may throw out a 
sufficient part of metals because of high velocity shock waves. But on the other the escape velocity 
is quite high for such galaxies and metals may be confined inside them. Whereas during star formation
burst dwarf galaxies can expel major part of their metal products in the IGM \citep{mix00}.
The velocity of shock waves produced by dwarfs is about several dozens km~s$^{-1}$.
% , but for massive galaxies it is hundreds km~s$^{-1}$. 
Low-velocity shock waves do not lead to the radiative 
precursor formation \citep{sd96}. The ionizing flux produced by the stellar population of a parent galaxy
is expected to be insignificant due to the burst character of star formation process (the massive 
part of the stellar population explodes as supernovae in a short timescale, whereas low mass 
stars do not produce sufficient number of ionizing photons). Therefore, the ionization and thermal 
evolution of gas behind low-velocity shocks is mainly governed by collisions between atoms, ions 
and electrons. 

The dynamics of interstellar/intergalactic gas in general may be understood by studying the ionization 
states of metals. Indeed, the CIV, NV, and OVI ions are sensitive tracers of hot gas with $T$ about 
several times $10^5$~K \citep[e.g.,][]{edgar}. Using theoretical models \citet{shull04} 
have investigated the ionization ratios of the Li-like absorbers CIV, NV and OVI in the Galactic halo,
\citet{gs09} have considered the ionic column densities as tracers of the thermal and ionization 
evolution behind strong radiative shocks with velocities more than hundred km~s$^{-1}$. Shock waves with 
different velocities can be identified using the spectral lines \citep{coxspec}. The growth of thermal 
instabilities can be detected by the associated X-ray and optical-ultraviolet lines \citep{bregman}. 
The analysis of the observed column densities of ions can give information about interstellar and 
intergalactic structures \citep[e.g.,][]{hvc1,simcoe06,gs07,agafonova,v10}. However, significant 
difficulties are connected to adequate interpretation of the obsevational data. Theoretical models 
of the ionization mechanisms give different predictions for ions and their ratios \citep[e.g.,][]{spitzer96}, 
and usually the ionic ratios do not remove the ambiguities regarding the ionization conditions 
\citep[e.g.,][]{shull04b}. But we can expect that ionizaton states of metals may help to recognize 
shock waves with $v \sim 50-150$~km~s$^{-1}$ and determine physical conditions in postshock gas. 

In this paper we study thermal instability in a collisionally controlled gas behind shock waves 
with $v \sim 50-150$~km~s$^{-1}$. A possible influence from external ionizing radiation field 
will be considered elsewhere.
The paper is organized as follows. In Section 2 we briefly describe the details of the model.
In Sections 3 and 4 we present our results. In Section 5 we summarize our results.

%----------------------- Section 2 -------------------------------

\section{Model description}

\noindent

Here we briefly describe our method of calculation. The full description of the method and the 
references to the atomic data can be found in \citet{v11}. We study the ionization and thermal 
evolution of a lagrangian element of cooling gas. In our calculations we consider all ionization 
states of the elements H, He, C, N, O, Ne, Mg, Si and Fe. We take into account the following major 
processes in a collisinal gas: collisional ionization, radiative and dielectronic recombination
as well as charge transfer in collisions with hydrogen and helium atoms and ions. 

The system of time-dependent ionization state equations should be complemented by the 
temperature equation. Neglecting the change of number of particles in the system (for 
fully ionized hydrogen and helium it remains approximately constant) the gas temperature 
is determined by 
\be
 {dT \over dt} = -{ n_e n_H  \Lambda \over A n k_B}
\ee
where $n$, $n_e$ and $n_H$ are the electron and total hydrogen number 
densities, $\Lambda(x_i,T,Z)$ is cooling rate, $A$ is a constant equal 
to $3/2$ for isochoric and $5/2$ for isobaric 
cooling, $k_B$ is the Boltzman constant. Cooling is isobaric when the 
cooling time is much greater then the dynamical time of a system, $t_c/t_d \gg 1$ 
and it is isochoric when the time ratio is opposite, $t_c/t_d \ll 1$. 

The total cooling rate is calculated using the photoionization code CLOUDY 
\citep[ver. 08.00,][]{cloudy}. For the solar metallicity we adopt the abundances reported 
by \citet{asplund}, except Ne for which the enhanced abundance is adopted \citep{drake}. 
In all our calculations we assume the helium mass fraction $Y_{\rm He} = 0.24$.
We solve a set of 96 coupled equations (95 for ionization states and one for temperature)
using a Variable-coefficient Ordinary Differential Equation solver \citep{dvode}. 

We consider pure collisional ionization model, so we should constrain shock wave velocity
by a value that does not lead to the radiative precursor formation. The precusor is photoionized by 
the radiation field emitted by a shocked gas, and consequently the initial ionization states in gas 
behind a shock front depends on the spectrum emitted by a shock wave. A stable photoionization 
precursor will be formed when the ionization front velocity in the gas approaching the shock 
becomes larger than the shock velocity. This condition is satisfied for shock velocity higher
than $v_*\simgt 175$~km~s$^{-1}$ \citep{sd96} or shock temperature 
$T_* = 3 m_p v_*^2/16 k \simgt 7\times 10^5$~K. Thus, we study the shock waves with temperature 
$T_s \simlt T_*$. We constrain our calculations by lower value $v = 155$~km~s$^{-1}$ corresponded 
to $T_s \simeq 6\times 10^5$~K. We assume that the gas temperature just behind a shock front 
instantaneously becomes equal to $T_s$, but the ionization composition relaxes from that corresponded to 
temperature $T_i$ before the front. During the relaxation period a gas cools more efficiently 
due to higher collisional ionization rates of low ionic states existed in the postshock gas, and 
the ionization composition tends to fit such thermal evolution. In general, for fixed other paramters
(density, metallicity) the ionization composition tends to that at a corresponding temeperature in 
a gas collisionally cooled from $T=10^8$~K. However, depending on the relation between $T_i$ and $T_s$ 
the relaxation period needs different time. 

\begin{figure}
\includegraphics[width=80mm]{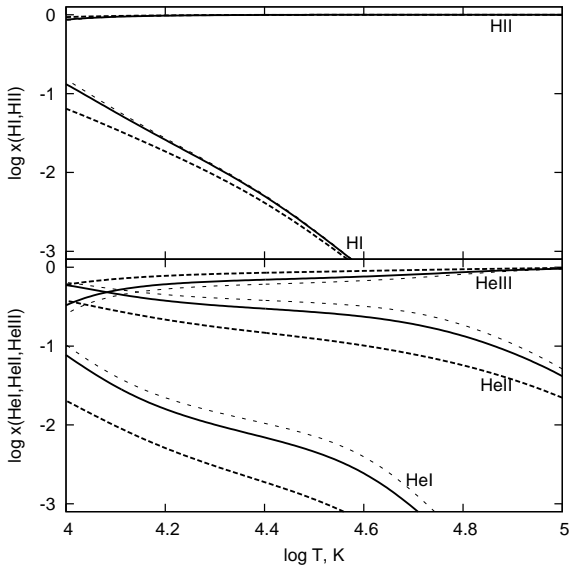}
\caption{
The hydrogen and helium ionization states for a gas isochorically cooled with $10^{-3}$, 0.1 and solar 
metallicities are shown by dot, solid and dash lines, respectively.
}
\label{figtihhe}
\end{figure}

The ionization times of H and He are longer than these of metals, so that the influence of H and He 
on the thermal evolution of gas just behind a shock is expected to be dominant. As an example 
Figure~\ref{figtihhe} presents the H and He ionization states in a gas isochorically cooled from 
$T=10^8$~K for $10^{-3}$, 0.1 and solar metallicities.
There is no significant dependence of the hydrogen states on metallicty, in contrast to that the helium 
states show remarkable distinction. The considerable dependence of the H and He states on temperature is 
clearly seen. Thus, taking different ionization composition corresponded $T=T_i$ in a gas before shock wave 
one can expect differences in the thermal evolution of the postshock gas. But the closer $T_i$ is to $T_s$, 
the smaller distinction of the thermal evolution is from that for a gas collisionally cooled from $T=10^8$~K. 
%% However, the increase of $T_i$ means that a shock wave becomes weaker.

A gas temperature before shock wave can vary in wide range. For example, a gas can be relaxed from some previous 
shock interaction or heated by galactic or extragalactic ionizing radiation. So it is difficult to choose
more or less common initial temperature. The gaseous temperature in HII regions is $\sim 2-4 \times 10^4$~K,
the typical temperature of the IGM is $\sim 2 \times 10^4$~K. To coinstrain calculations we suppose that the 
initial temperature of a gas before shock wave equals $T_i=2\times 10^4$~K, and a gas has the ionization 
composition obtained in the non-equilibrium (time-dependent) model at this temperature and given metallicity. 
But in several sets of models we study the depedence of the gas evolution on $T_i < T_s$. For these models the 
ionization composition is corresponded to $T_i$. 

%%%%%%%%%%%%%%%%%%%%%%%%%%%%%%%%%%%%%%%%%%%%%%%%%%%%%%
%%%%%%%%%%%%%%%%%%%%%%%%%%%%%%%%%%%%%%%%%%%%%%%%%%%%%%

%----------------------- Section 3 -------------------------------

\section{Cooling rates and thermal instability}

%%%%%%%%%%%%%%%%%%%%%%%%%%%%%%%%%%%%%%%%%%%%%%%%%%%%%%
\begin{figure}
\includegraphics[width=80mm]{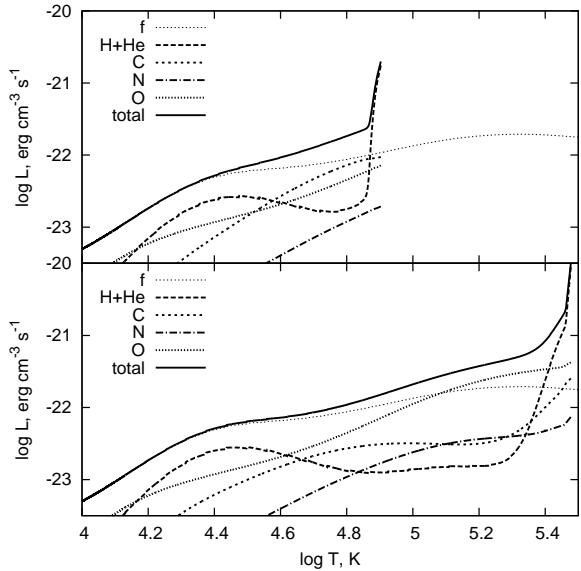}
\caption{
The contributions to the total cooling rate from each chemical element for $T_i = 8\times 10^4$~K
(upper panel) and $T_i = 3\times 10^5$~K (lower panel) in a gas with solar metallicity. The cooling
rate for the fiducial model, $T_i = 10^8$~K, is shown by thin dotted line.
}
\label{figticc}
\end{figure}
%%%%%%%%%%%%%%%%%%%%%%%%%%%%%%%%%%%%%%%%%%%%%%%%%%%%%%

The thermal instability criteria for a gas cooled isochorically and isobarically are \citep{yua78}
\be
f = {\rm d~ln~\Lambda \over d~ln~T} < 1 , f = {\rm d~ln~\Lambda \over d~ln~T} < 2,
\ee
respectively. The isochoric and isobaric cooling rates for a gas cooled from $T=10^8$~K are close to each 
other \citep{gs07}. So that a gas is unstable to isochoric/isobaric perturbations in very 
close temperature ranges. However, the cooling rate of a gas behind a shock may depend on physical 
condition in a gas before shock front, e.g. ionization composition and temperature of a gas, so that 
the temperature range, where a gas becomes thermally unstable, can be modified. 

\begin{figure*}
\includegraphics[width=160mm]{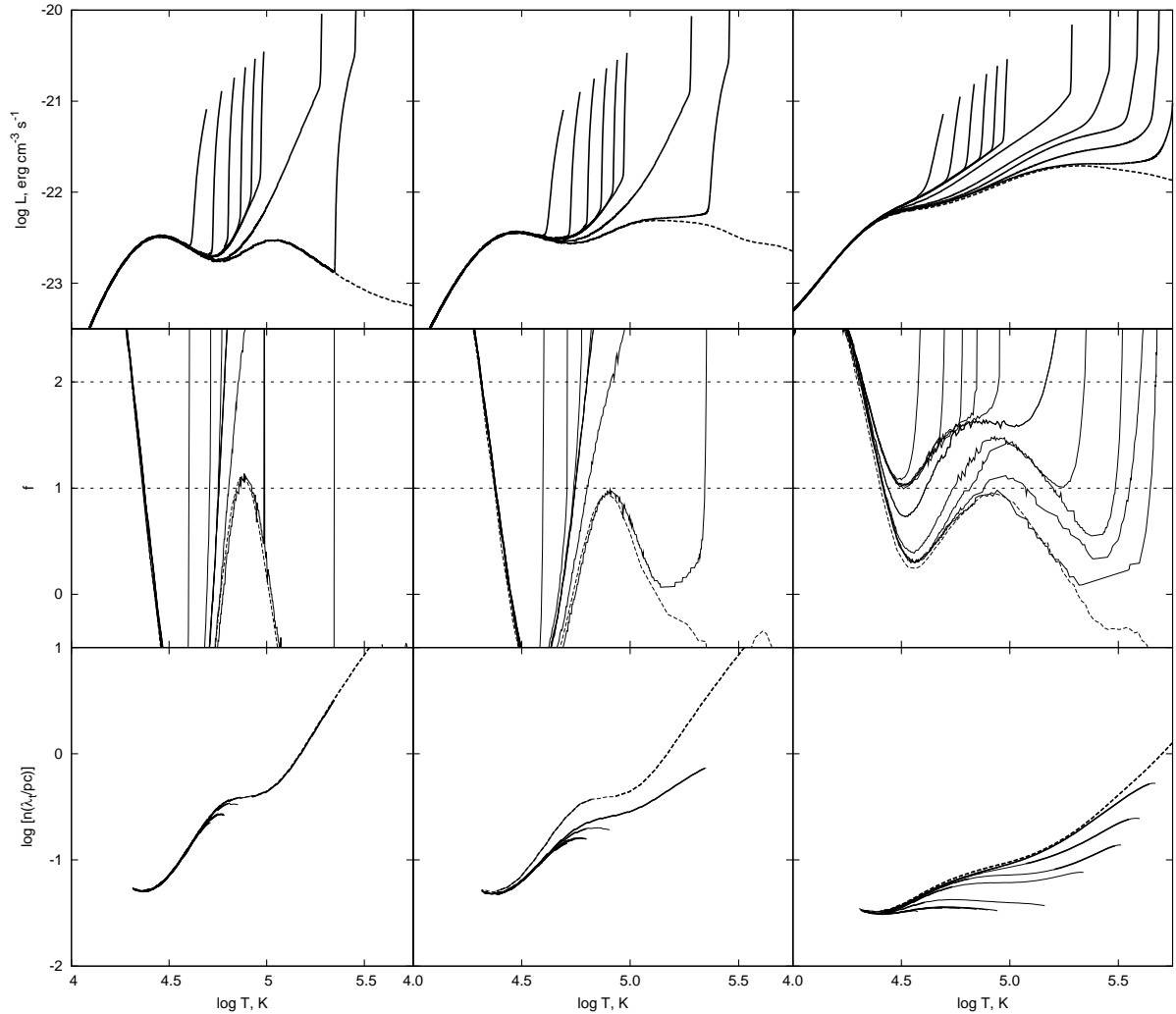}
\caption{
The cooling rates (upper panels), thermal instability criterion (middle) and characteristic 
scale (lower) for gas wih $10^{-3}$, 0.1 and 1 $Z_\odot$ metallicities (from left to right). 
{Upper panels.} The cooling rates for $T_s = 5\times 10^4$, $6\times 10^4$, $7\times 10^4$, $8\times 10^4$, 
$9\times 10^4$, $10^5$, $2\times 10^5$, $3\times 10^5$~K are depicted by solid lines from left to right, 
respectively. For solar metallicity (right column of panels) the cooling rates for $T_s = 4\times 10^5$, 
$5\times 10^6$ and $6\times 10^5$~K are added. The cooling rate for the fiducial model is shown by dash line. 
{Middle panels.} The same for the thermal instability criterion. Two horizontal lines correspod to the
criterion for isobaric (upper) and isochoric (lower line) perturbations. A gas is unstable below the lines.
{Lower panels.} The column density of a thermally unstable gas layer, $N_t = n/\lambda$, for isobaric 
(shown by thin solid lines) and isochoric (shown by thick solid lines) perturbations, where $\lambda_t = c_s t_{cool}$ 
is the thermal length. The lines from bottom to top correspond to higher shock temperature, $T_s$.
For $Z=10^{-3}Z_\odot$ the lines coincide.
}
\label{figtic}
\end{figure*}

Figure~\ref{figtic} presents the isochoric cooling rates, thermal instability criterion, $f$, and 
characteristic scale for a gas with $10^{-3}$, 0.1 and 1 $Z_\odot$ behind a shock with $T_s$. Below we 
present our calculations in terms of a shock temperature $T_s = 3 m_p v_s^2/16 k \simeq 2.3\times 10^3v_{10}^2$, 
where $v_{10} = v_s/10$~km~s$^{-1}$.
Also for a given metallicity the isochoric cooling rate for $T_i=T_s=10^8$~K, a {\it fiducial} model, 
which corresponds to the tabulated cooling rates \citep[e.g.,][]{sd93,gs07}, is plotted by thin solid line. 
First of all we should note that no significant difference between isobaric and isochoric cooling rates for
$T_s\le 6\times 10^5$~K is found. So here we present figures for the isochoric case only, but due to the
different thermal instability criterion we analyze both cases. 

%%%%%%%%%%%%%%%%%%%%%%%%%%%%%%%%%%%%%%%%%%%%%%%%%%%%%%
%%%%%%%%%%%%%%%%%%%%%%%%%%%%%%%%%%%%%%%%%%%%%%%%%%%%%%

Figure~\ref{figtic}(upper panels) shows that the cooling rates for the whole shock temperature range considered 
here, $T_s = 0.5 - 6\times 10^5$~K, tend to the fiducial one and equal to it at low temperature. For gas with 
$Z=10^{-3}~Z_\odot$ this occurs almost at the $T_s$ due to fast ionization of hydrogen for $T_s \simlt 7\times 10^4$~K 
and both H and He ionization for higher initial temperature (the vertical part of lines in panel a). For higher 
metallicity other chemical elements, mainly carbon and oxygen, can dominate 
in cooling. This is clearly seen in Figure~\ref{figticc}, which presents the contributions to the total cooling 
rate from each chemical element for $T_s = 8\times 10^4$~K (upper panel) and $T_s = 3\times 10^5$~K (lower panel) 
in a gas with solar metallicity. One note that the difference between cooling rates behind shock waves and the 
rate for the fiducial model reaches maximum for solar metallicity and depends strongly on $T_s$, but for any 
$T_s$ the cooling rates coincide at $T \simlt 3\times 10^4$~K. Thus, we can conclude that the cooling rate in a 
gas behind a shock wave differs considerably from the cooling rate for the fiducial model, $T_s = 10^8$~K, which 
is usually used to study the thermal evolution of a gas.

Panels (b) present the $f$ value for the considered models. Two horizontal lines correspond to the isobaric/isochoric
thermal instability criteria. Note that both criteria are satisfied for the fiducial model at $T\simgt 3\times 10^4$~K
for any metallicity considered here. For $Z=10^{-3}~Z_\odot$ both isobaric and isochoric criteria for the models
with $T_s = 0.5 - 6\times10^5$~K are also satisfied (the region below horizontal lines) in the same temperature range 
as that for the fiducial model, except for very early evolution of a gas behind shock wave (almost vertical tails 
of the cooling curves correpond to that period of the evolution). Increase of metallicity leads to widening 
temperature range, where the isochoric criterion is not satisfied, but does not change the range for the isobaric 
case. A gas evolved from $T_s = 0.5-6\times 10^5$~K is thermally unstable to isobaric perturbations during almost
full its evolution. In a gas cooled isochorically the value $f$ for the solar metallicity demonstrates a significant 
difference from that for the fiducial model. The temperature range, where a gas is unstable behind
shock waves with $T_s\simgt 6\times 10^5$~K, is very close to that for the fiducial model. But such temperature 
range becomes narrower for lower $T_s$, e.g. for $T_s = 3\times 10^5$~K a gas is unstable at
$T\sim 2.6-5.4\times 10^4$~K, and for $T_s = 10^5$~K it proceeds at $T\sim 2.8-4\times 10^4$~K. For 
$T_s \simlt 10^5$~K a gas remains thermally stable to isochoric perturbations during full its evolution. 

%%%%%%%%%%%%%%%%%%%%%%%%%%%%%%%%%%%%%%%%%%%%%%%%%%%%%%
\begin{figure}
\includegraphics[width=70mm]{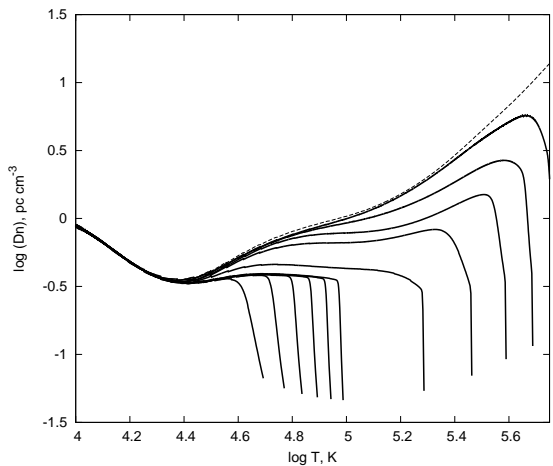}
\caption{
The critical column density, $D_{tr}n$, for the models with $T_s = 5\times 10^4$, $6\times 10^4$, $7\times 10^4$, $8\times 10^4$, $9\times 10^4$, $10^5$, $2\times 10^5$, $3\times 10^5$, $4\times 10^5$, $5\times 10^6$ and 
$6\times 10^5$~K (thick dashed lines from left to right, respectively) and for the fiducicial model (thin solid 
line) at solar metallicity.
}
\label{figtitr}
\end{figure}
%%%%%%%%%%%%%%%%%%%%%%%%%%%%%%%%%%%%%%%%%%%%%%%%%%%%%%

%%%%%%%%%%%%%%%%%%%%%%%%%%%%%%%%%%%%%%%%%%%%%%%%%%%%%%
\begin{figure}
\includegraphics[width=70mm]{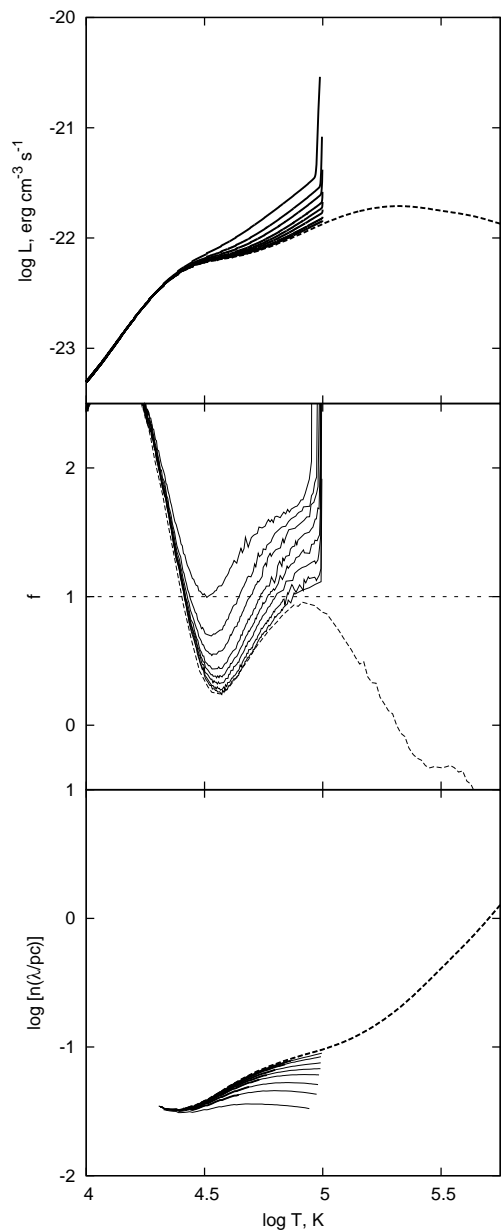}
\caption{
The same as in Figure~\ref{figtic}, but for the shock wave with $T_s = 10^5$~K depending on the initial 
ionization composition, which is corresponded to $T_i = 2 \times 10^4$, $3 \times 10^4$, $4 \times 10^4$, 
$5 \times 10^4$, $6 \times 10^4$, $7 \times 10^4$, $8 \times 10^4$ , $9 \times 10^4$~K from top to bottom,
respectively.
}
\label{figticvsT}
\end{figure}
%%%%%%%%%%%%%%%%%%%%%%%%%%%%%%%%%%%%%%%%%%%%%%%%%%%%%%

Panels (c) present the column density of a gas layer, which can be thermally unstable, $N_t = n/\lambda_t$,
where $\lambda_t = c_s t_{cool}$ is the thermal length. We have plotted the dependence of $N_t$, where the instability 
criterion is satisfied, namely, for any $T_s$ at $T\simgt 2\times 10^4$~K in the isobaric case (shown by thin
solid lines) as well as for $T_s\simgt 10^5$~K in the isochoric one (shown by thick solid lines). Note that 
the criterion for the fiducial model with $T_i=10^8$~K is satisfied for 
$T\simgt 2\times 10^4$~K (dash line). The difference between column densitites for $T_s = 0.5-3\times 10^5$~K 
and these for the fiducial model reaches a factor of 2--3 for $Z\simgt 0.1$, it is maximum for early stages 
of the evolution, where the cooling rate is changed considerably.

Following Gnat \& Sternberg (2007) we consider the transition from isobaric to isochoric cooling for the
interaction a shock wave with a cloud in terms of the critical column density, $D_{tr}n$, and temperature, 
where $D_{tr}$ is a critical size of cloud. Figure~\ref{figtitr} present such dependence for the set of 
models with $T_s = 0.5 - 6\times 10^5$~K (solid lines) and for the fiducial model (dash line) for solar 
metallicity. The cooling becomes isochoric for temperatures and cloud column densities left and above 
of the curves. Here $D_{tr}n \propto c_s$, so that $D_{tr}n$ should be multiplied by the Mach number $M$ 
for a shock wave with $v_s = M c_s$. 

In the above-considered set of models we start from the ionization composition corresponded to the
$T_i = 2\times 10^4$~K. But a gas behind shock wave can have another ionization composition. Figure~\ref{figticvsT} 
presents the isochoric cooling rates, thermal instability criterion and characteristic scale for $T_s = 10^5$~K
depending on the initial ionization composition, which is corresponded to $T_i = (2 - 9) \times 10^4$~K. 
Figure~\ref{figticvsT}(upper panel) shows that the cooling rates gradually approach to the fiducial one, 
the difference is negligible for $T_i \simgt 7 \times 10^4$~K. The thermal instability criterion and the value 
$n\lambda_t$ also demonstrate the same behavior.

%----------------------- Section 4 -------------------------------

\section{Ionization states and their ratios}

%%%%%%%%%%%%%%%%%%%%%%%%%%%%%%%%%%%%%%%%%%%%%%%%%%%%%%
\begin{figure}
\includegraphics[width=80mm]{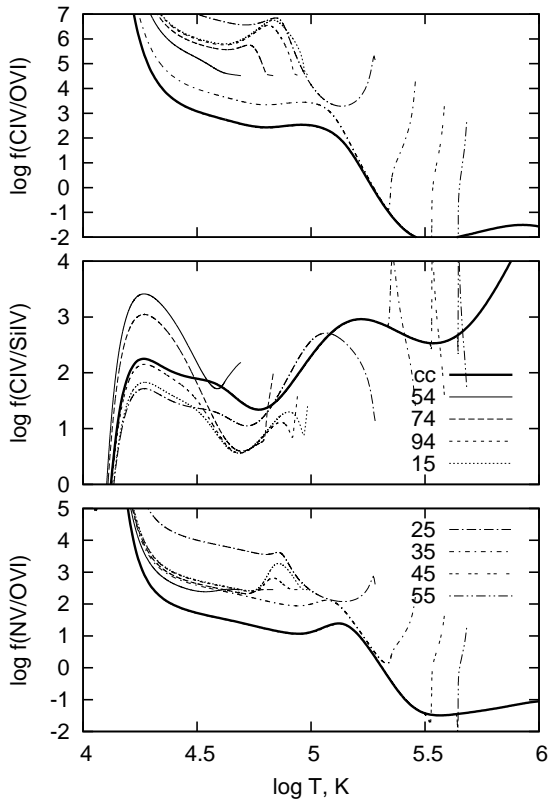}
\caption{
The $N_{\rm NV}/N_{\rm OVI}$, $N_{\rm CIV}/N_{\rm SiIV}$ and $N_{\rm CIV}/N_{\rm OVI}$ ionization 
ratios for $10^{-3}Z_\odot$ metallicity gas with $T_i = 2\times 10^4$~K behind a shock 
wave with $T_s = 5\times 10^4$~K (thin solid line), $7\times 10^4$~K (dash line),  $9\times 10^4$~K 
(short dash line),  $10^5$~K (dot line),  $2\times 10^5$~K (dash-dot line),  $3\times 10^5$~K (short 
dash-dot line), $4\times 10^5$~K (short dash-dash line), $5\times 10^5$~K (dash-dot-dot line) and 
the fiducial model, $T_s = 10^8$~K, (thick solid line).
}
\label{figtinvz3}
\end{figure}
%%%%%%%%%%%%%%%%%%%%%%%%%%%%%%%%%%%%%%%%%%%%%%%%%%%%%%

%%%%%%%%%%%%%%%%%%%%%%%%%%%%%%%%%%%%%%%%%%%%%%%%%%%%%%
\begin{figure}
\includegraphics[width=80mm]{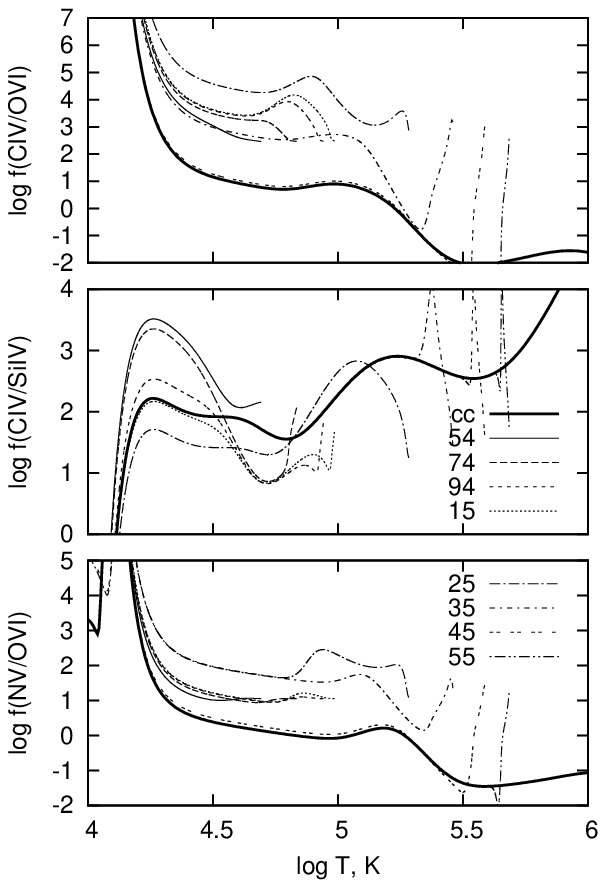}
\caption{
The same as in Figure~\ref{figtinvz3}, but for $Z=0.1Z_\odot$.
}
\label{figtinvz1}
\end{figure}
%%%%%%%%%%%%%%%%%%%%%%%%%%%%%%%%%%%%%%%%%%%%%%%%%%%%%%

%%%%%%%%%%%%%%%%%%%%%%%%%%%%%%%%%%%%%%%%%%%%%%%%%%%%%%
\begin{figure}
\includegraphics[width=80mm]{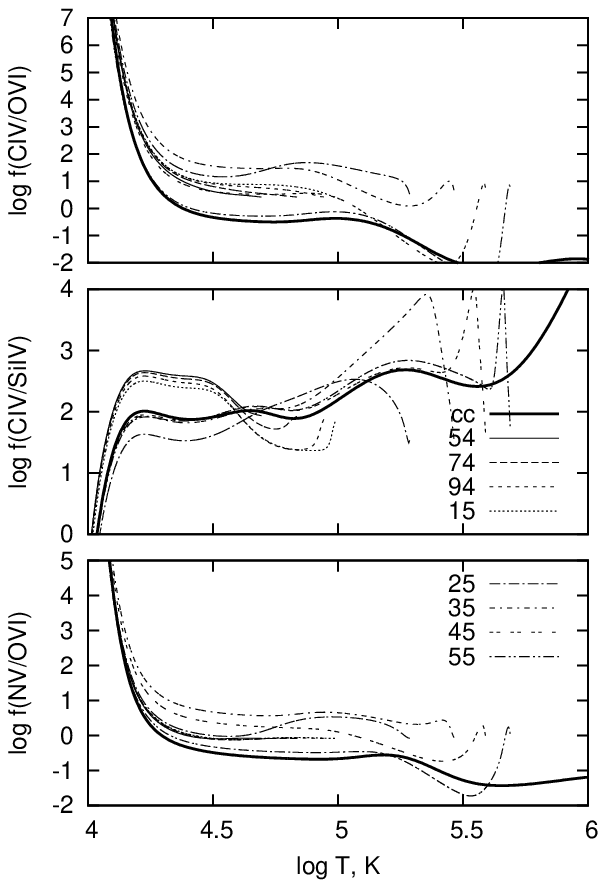}
\caption{
The same as in Figure~\ref{figtinvz3}, but for $Z=Z_\odot$.
}
\label{figtinvzs}
\end{figure}
%%%%%%%%%%%%%%%%%%%%%%%%%%%%%%%%%%%%%%%%%%%%%%%%%%%%%%

%%%%%%%%%%%%%%%%%%%%%%%%%%%%%%%%%%%%%%%%%%%%%%%%%%%%%%
\begin{figure}
\includegraphics[width=80mm]{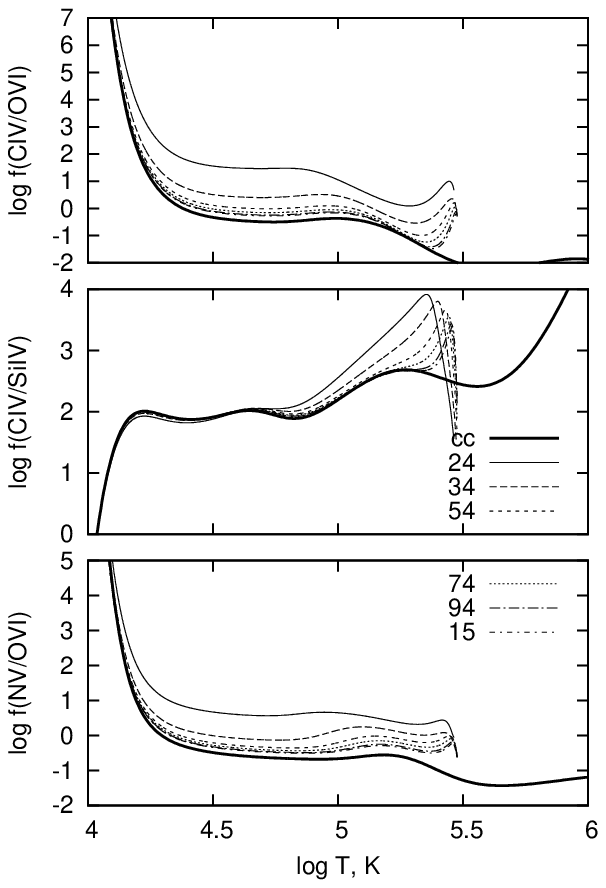}
\caption{
The $N_{\rm NV}/N_{\rm OVI}$, $N_{\rm CIV}/N_{\rm SiIV}$ and $N_{\rm CIV}/N_{\rm OVI}$ ionization 
ratios for solar metallicity gas with $T_i = 2\times 10^4$~K (thin solid line), 
$3\times 10^4$~K (dash line),  $5\times 10^4$~K (short dash line),  $7\times 10^4$~K (dot line),  
$9\times 10^4$~K (dash-dot line),  $10^5$~K (short dash-dot line) behind a shock wave with 
$T_s = 3\times 10^5$~K. The ratios for the fiducial model, $T_s = 10^8$~K, are depicted by thick solid line).
}
\label{figtinvzs35}
\end{figure}
%%%%%%%%%%%%%%%%%%%%%%%%%%%%%%%%%%%%%%%%%%%%%%%%%%%%%%

Figures~\ref{figtinvz3}-\ref{figtinvzs} show the $N_{\rm NV}/N_{\rm OVI}$, $N_{\rm CIV}/N_{\rm SiIV}$ and 
$N_{\rm CIV}/N_{\rm OVI}$ ionization ratios in a gas with $10^{-3}$, 0.1 and 1$Z_\odot$ metallicities behind 
a shock wave with $T_s$. The temperature before a shock wave is taken $T_i = 2\times 10^4$~K. Here we consider 
isochoric ionization ratios, although the difference from isobaric ratios is non-negligible, their temperature 
dependences show similar behavior. 

The ionization ratios for the metallicities considered here demonstrate a significant dependence on $T_s$ and difference from the ratios in the fiducial model. The difference in some temperature ranges reaches several orders of 
magnitude. The difference of the $N_{\rm CIV}/N_{\rm OVI}$ ratio increases with the growth of $T_s$ and reach 
maximum (up to two orders of magnitude) for $T_s = 2-3\times 10^5$~K in a gas with $Z=10^{-3} - 1Z_\odot$ 
(see the lower panels in Figures~\ref{figtinvz3}-\ref{figtinvzs}). Further increase of $T_s$ diminishes the 
difference rapidly, and the ratios for $T_s = 4-5\times 10^5$~K almost coincide with the fiducial one. Stronger 
shock wave (higher $T_s$) leads to the ionization of higher states, and the difference from the fiducial model 
should decrease with the growth $T_s$. Two other ratios exhibit similar behavior.

The difference of the ionic ratios from the fiducial ones is maximum for the lowest metallicity, $Z=10^{-3}Z_\odot$. 
For example, the $N_{\rm CIV}/N_{\rm OVI}$ ratio for $T_s = 2\times 10^5$~K  differs about four orders of magnitude 
from that in the fiducial model. In a solar metallicity gas this difference reaches two orders only. 
Obviously, the ionic fractions decrease fastly in low metallicity gas \citep[e.g.][]{gs07}, and the recombination 
lag increases with the metallicity growth. So the comparison of two ions with large gap between ionization
potentials, like CIV and OVI (47.9 and 113.9~eV for CIII$\rightarrow$CIV and OV$\rightarrow$OVI, respectively), 
gives higher values for lower metallicity. Whereas the ratio between CIV and SiIV demonstrate weaker dependence 
on metallicity due to proximity of their ionization potentials (34~eV for SiIII$\rightarrow$SiIV). 

Such a strong dependence on $T_s$ may lead to some uncertainity or inaccuracy of determining the physical 
conditions in the postshock gas using the tabulated collisional ionization states \citep[e.g.][]{sd93,gs07}. 
In the solar metallicity gas the most remarkable dependence is within the temperature range, where the postshock 
gas can be thermally stable for isochoric perturbations (see Figure~\ref{figtic}) and the transition from isobaric 
to isochoric cooling is more probable (Figure~\ref{figtitr}). Moreover, the postshock material can be efficiently 
mixed due to the hydrodynamical instabilities \citep{slavin,avillez09}, and a degree of uncertainity of determining 
the physical conditions may increase. So that the strong dependence of ionic states can hardly help to dicriminate 
this transition, and it should be taken into account for correct interpretation of observational data and synthetic 
data models obtained from numerical simulations. 

Also we have analyzed the dependence on the initial ionization composition of a gas. Figure~\ref{figtinvzs35} shows 
the same ionization ratios, $N_{\rm NV}/N_{\rm OVI}$, $N_{\rm CIV}/N_{\rm SiIV}$ and $N_{\rm CIV}/N_{\rm OVI}$, in 
a gas behind a shock wave with $T_s = 3\times 10^5$~K for different initial ionization composition corresponded to 
$T_i$. The increase of $T_i$ leads to less difference of the ionic ratios from these in the fiducial model. As it 
is mentioned above such difference is more pronounced for ions with larger deviation in their ionization potentials. 

%%%%%%%%%%%%%%%%%%%%%%%%%%%%%%%%%%%%%%%%%%%%%%%%%%%%%%
\begin{figure}
\includegraphics[width=100mm]{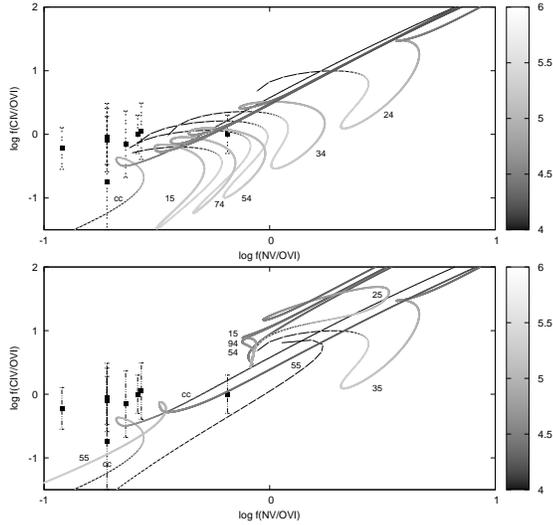}
\caption{
{\it Lower panel}. The dependence $N_{\rm CIV}/N_{\rm OVI}$ versus $N_{\rm NV}/N_{\rm OVI}$  
for solar metallicity gas with $T_i = 2\times 10^4$~K behind a shock wave with $T_s = 5\times 10^4$~K (label '54'), 
$9\times 10^4$~K (label '94'),  $10^5$~K (label '15'),  $2\times 10^5$~K (label 25), $3\times 10^5$~K 
(label '35') and the fiducial model, $T_s = 10^8$~K, (label 'cc').
{\it Upper panel}. The dependence $N_{\rm CIV}/N_{\rm OVI}$ versus $N_{\rm NV}/N_{\rm OVI}$ 
for solar metallicity gas with $T_i = 2\times 10^4$~K (label '24'), $3\times 10^4$~K (label '34'), 
$5\times 10^4$~K (label '54'),
$7\times 10^4$~K (label '74'), $9\times 10^4$~K (label '94'),  $10^5$~K (label '15') behind a shock wave with 
$T_s = 3\times 10^5$~K. The ratios for the fiducial model, $T_s = 10^8$~K, are marked by label 'cc'.
Gas temperature is indicated by gray scale along the trajectories: from hot (light gray) to cold (black) gas.
The data points show the ionic ratios observed in metal absorbers \citep[see Table 8][]{hvc1}.
}
\label{figtiratio}
\end{figure}

Figure~\ref{figtiratio} (lower panel) presents the dependence of $N_{\rm CIV}/N_{\rm OVI}$ versus 
$N_{\rm NV}/N_{\rm OVI}$ column densities on the temperature behind a shock wave with $T_s$ for a gas 
with solar metallicity and the initial ionization composition corresponded to $T_i = 2\times 10^4$~K.
All tracks start from the same point corresponded to the initial ratio at $T_i = 2\times 10^4$~K. Further 
evolution depends on collisional ionization rate of ions behind a shock with $T_s$. As it is expected
from the above-mentioned results the tracks differ significantly from the fiducial model: the largest 
deviation can be found for $T_s \sim 2-3\times 10^5$~K, while the tracks for $T_s \simlt 10^5$~K are 
close to each other, and the tracks for $T_s \simgt 4\times 10^5$~K becomes closer
to that for the fiducial model. The difference becomes negligible at $T_s\simgt 6\times 10^5$~K. 

Finally we again consider the dependence on the initial ionization composition of a gas. 
Figure~\ref{figtiratio} (upper panel) shows the dependence of the column densities behind a shock wave with
$T_s = 3\times 10^5$~K on the initial ionization composition corresponded to $T_i$. We should 
note that the tracks for the shock models tends gradually to the fiducial track, but these tracks coincide at 
$T\simlt 4\times 10^4$~K. However, at $T\simgt 4\times 10^4$~K a significant difference can be found even for 
the track $T_i = 10^5$~K. So that even small deviation of the initial ionic composition leads to remarkably 
different tracks. 
In both panels we add the observational data for the high velocity clouds in the Galactic halo 
\citep[see Table 8 in][]{hvc1}. Although several tracks are within the errorbars of the observational 
data, the majority of the points cannot be fitted by any track presented in Figure~\ref{figtiratio}.
These observational points are belived to be explained by turbulent mixing behind shock waves 
\citep{slavin,avillez09}.

%----------------------- Section 5 -------------------------------
\section{Conclusions}

\noindent

In this paper we have studied the thermal instability, nonequilibrium cooling rates and ionization states 
in a collisionally controlled gas behind shock waves with $v \sim 50-150$~km~s$^{-1}$ ($T_s \sim 0.5 - 
6\times10^5$~K). Such shock waves do not lead to the radiative precursor formation, and gas evolution is
governed by collisions only. 

Our results can be summarized as follows.
\begin{itemize}
 \item The cooling rate in a gas behind shock waves with $T_s \sim 0.5 - 6\times 10^5$~K differs 
  considerably from the cooling rate for a gas cooled from $T = 10^8$~K.
 \item The temperature range, where the postshock gas is thermally unstable, is significantly modified 
  and depends on gas metallicity. For $Z \sim 10^{-3}~Z_\odot$ both isobaric and isochoric criteria for 
  shock waves with $T_s = 0.5 - 6\times10^5$~K and $T_s = 10^8$~K are satisfied in the same temperature 
  range. Increase of metallicity leads to widening temperature range, where the criterion is not fulfilled. 
  For the solar metallicity a gas behind a shock with $v \sim 50-120$~km~s$^{-1}$ ($T_s = 0.5-3\times 10^5$~K) 
  is thermally unstable to isobaric perturbations during almost full its evolution. But the temperature range, 
  where a gas is unstable to isochoric perturbations, becomes narrower for lower $T_s$, and a gas remains 
  thermally stable 
  to isochoric perturbations behind shock waves with $v \simlt 65$~km~s$^{-1}$ ($T_s \simlt 10^5$~K) during full 
  its evolution. 
 \item The column density of a gas layer, which can be thermally unstable, also depends on gas metallicity.
  The difference between column densitites for $T_s = 0.5-3\times 10^5$~K and these for a gas cooled from 
  $T = 10^8$~K reaches a factor of 2--3 for $Z\simgt 0.1Z_\odot$.
 \item The transition from isobaric to isochoric cooling for a gas with $Z \simgt 0.1~Z_\odot$ behind shock 
  waves with $T_s = 0.5 - 3\times10^5$~K proceeds earlier (at lower column density layer) than that occurs in a 
  gas cooled from $T = 10^8$~K.
 \item The ionic ratios in a gas with $Z \sim 10^{-3} - 1Z_\odot$ behind shock waves with $T_s \simlt 4\times10^5$~K 
  demonstrate a significant dependence on $T_s$ and difference from the ratios in a gas cooled from $T = 10^8$~K. 
  The difference becomes negligible at $T_s\simgt 6\times 10^5$~K in the metallicity range considered here. 
\end{itemize}

%----------------------- Section K -------------------------------
\section{Acknowledgements}

\noindent

The author is grateful to Yuri Shchekinov for help and many useful discussions. 
Gary Ferland and CLOUDY community are acknowledged for creating of the excellent tool
for study of the photoionized plasma -- CLOUDY code.
This work is supported by the RFBR (project codes 09-02-00933, 11-02-90701),
by the Federal Agency of Education (project code RNP 2.1.1/1937)
and by the Federal Agency of Science and Innovations (project 02.740.11.0247). 
%% The author acknowledges the Institute of Astronomy RAS for hospitality, where he was 
%% a visiting scientist.

%----------------------- Section L -------------------------------

\end{document}